\documentclass[12pt,preprint]{aastex}

\newcommand{\zh}{\mbox{[$Z$/H]}}
\newcommand{\alphafe}{\mbox{$[\alpha/{\rm Fe}]$}}
\newcommand{\vi}{\mbox{$V\!-\!I$}}
\newcommand{\vio}{\mbox{$(V\!-\!I)_0$}}
\newcommand{\vk}{\mbox{$V\!-\!K$}}
\newcommand{\fe}{\mbox{$\langle {\rm Fe} \rangle$}}
\newcommand{\mgfe}{$\mbox{[MgFe]}$}
\newcommand{\msun}{\mbox{M$_{\odot}$}}
\newcommand{\mgtwo}{\mbox{Mg$_2$}}
\begin{document}
\shorttitle{Globular clusters in NGC 4365}
\title{Evidence for An Intermediate-age, Metal-rich Population of 
       Globular Clusters in NGC~4365\footnote{
 Based on data obtained at the W.~M.~Keck Observatory, 
 which is operated as a scientific partnership among the California Institute 
 of Technology, the University of California and the National Aeronautics and
 Space Administration.}
}
\author{S{\o}ren S. Larsen and Jean P. Brodie  
  \affil{UCO / Lick Observatory, UC Santa Cruz, USA} 
  \email{soeren@ucolick.org, brodie@ucolick.org}
\and
  Michael A. Beasley and Duncan A. Forbes 
   \affil{Astrophysics \& Supercomputing, Swinburne University, 
          Hawthorn VIC 3122, Australia} 
   \email{dforbes@astro.swin.edu.au and mbeasley@astro.swin.edu.au}
\and
  Markus Kissler-Patig and Harald Kuntschner
  \affil{European Southern Observatory, Karl-Schwarzschild-Stra\ss e 2,
         D-85748 Garching b.\ M{\"u}nchen, Germany} 
  \email{mkissler@eso.org, hkuntsch@eso.org}
\and
  Thomas H. Puzia
  \affil{Sternwarte M{\"u}nchen, Scheinerstr.\ 1, D-81679 M{\"u}nchen, Germany}
  \email{puzia@usm.uni-muenchen.de}
}

\slugcomment{Accepted for publication in ApJ}

\begin{abstract}
  We present spectroscopy for globular clusters (GCs) in the elliptical galaxy 
NGC~4365, obtained with the LRIS spectrograph on the Keck I telescope.
Previous studies have shown that the optical color distribution of GCs in 
NGC~4365 lacks the bimodal structure that is common in globular cluster 
systems, showing only a single broad peak.  Measurements of Balmer line 
indices (H$\beta$, H$\gamma$ and H$\delta$) on the GC spectra support recent 
suggestions by Puzia et al., based on optical and near-infrared photometry, 
that some of the clusters in NGC~4365 are intermediate-age (2--5 Gyrs) and 
metal-rich ($-0.4\la\zh\la0$) rather than old ($\sim$10--15 Gyrs) and 
metal-poor. We also find some genuinely metal-poor, old clusters, suggesting 
that the ages and metallicities of the two populations conspire to produce the 
single broad distribution observed in optical colors.
\end{abstract}

\keywords{galaxies:elliptical and lenticular,cD ---
          galaxies:individual (NGC~4365) ---
          galaxies:evolution ---
	  globular clusters: general}

\section{Introduction}

  Several recent studies have shown that complicated color distributions in
globular cluster systems (GCSs) are the norm rather than the exception 
\citep{gk99,kw01,lar01}.  In many cases, the GCS color distributions can be 
fairly well modeled as a superposition of two Gaussians with peaks at 
$\vio\approx0.95$ and $\vio\approx1.18$. These colors correspond to mean 
metallicities of $\zh=-1.4$ and $\zh=-0.6$ \citep{kis98} if both populations 
are ``old'', i.e.\ they have similar ages (10--15 Gyrs) as those in 
the Milky Way.

  Although the color differences of globular clusters have generally been 
assumed to be mainly a function of metallicity, it is clear that age 
differences between GC sub-populations will also affect their colors, making 
a younger population appear bluer for a given metallicity.  There are a few 
cases in which only a single, broad peak is observed in the optical 
GC color distribution, usually at blue or intermediate colors.  While these 
systems might simply lack a distinct metal-rich component, which would 
normally show up as the red peak, another interesting possibility is that 
the metal-rich population is indeed present but has a significantly lower age 
than the metal-poor one, shifting it to bluer colors and making the two 
populations appear as a single broad peak in optical colors.  Using optical 
photometry alone, this degeneracy between age and metallicity cannot 
easily be broken and alternative methods are needed to provide independent 
constraints on metallicity or age (or both). 

  Combining optical and near-infrared imaging of GCs in the $\sim3$ Gyr old 
merger remnant NGC~1316, 
\citet{goud01} have recently demonstrated that an old, metal-poor population 
and a very metal-rich, younger population with an estimated age consistent 
with that of the merger \citep{kd98}, masquerade as a single peak in optical 
colors.  It is possible that other systems with an apparently unimodal GC 
color distribution might in a similar way be composed of several populations 
with large age differences.

  The giant elliptical galaxy NGC~4365 at the outskirts of the Virgo cluster 
is another example of a galaxy with only a single broad peak in the GC \vi\ 
color distribution \citep{for96,gk99,lar01}.  From optical and near-infrared 
photometry, \citet[][hereafter P02]{puz02} have suggested that some of the GCs 
in NGC~4365 might indeed be intermediate-age (2--8 Gyrs) and very metal-rich 
($Z_{\odot} - 3 Z_{\odot}$).  In this paper we use new spectroscopic data
to further investigate the ages and metallicities of GCs in NGC~4365.

\section{Data}
\label{sec:data}

  Spectra for GC candidates in NGC~4365 were obtained in multi-slit mode on 
2002 Feb.\ 9 and Feb.\ 10 with the LRIS spectrograph \citep{oke95} on the 
Keck I telescope.  Candidate young clusters were selected from the data in 
P02, but to fill up the slitmask a number of objects without $K$-band imaging 
were also included.  We obtained 9 exposures with integration times of 30--60 
min, yielding a total exposure time of 390 min (6$\frac{1}{2}$ hours). 
Observations were carried out simultaneously with the blue and red sides on 
LRIS, using a dichroic splitting at 5600\AA . On the blue side we used a 600 
l/mm grism, covering $\lambda\lambda$3800--5600\AA , while a 600 l/mm grating 
blazed at 5000\AA\ was used on the red side, covering the range 
$\lambda\lambda$5600--7900\AA .  A number of radial velocity and flux 
standards, as well as Lick/IDS standard stars from \citet{wor94}, were also 
observed. The radial velocity standards were picked
from the compilation in \citet{bar00} (numbers 9934 and 10927) while the 
flux standards were PG0934$+$554 and Hiltner 600 \citep{mas88}.
Because the spectral range covered in longslit mode is slightly different
from the one covered in multislit mode, we observed the flux standards 
in multislit mode through one of the slitlets in the NGC~4365 slitmask,
to facilitate flux calibration over the entire spectral range of the science 
spectra. A slit width of $1\farcs0$ was used for all observations.

\begin{figure}
\plotone{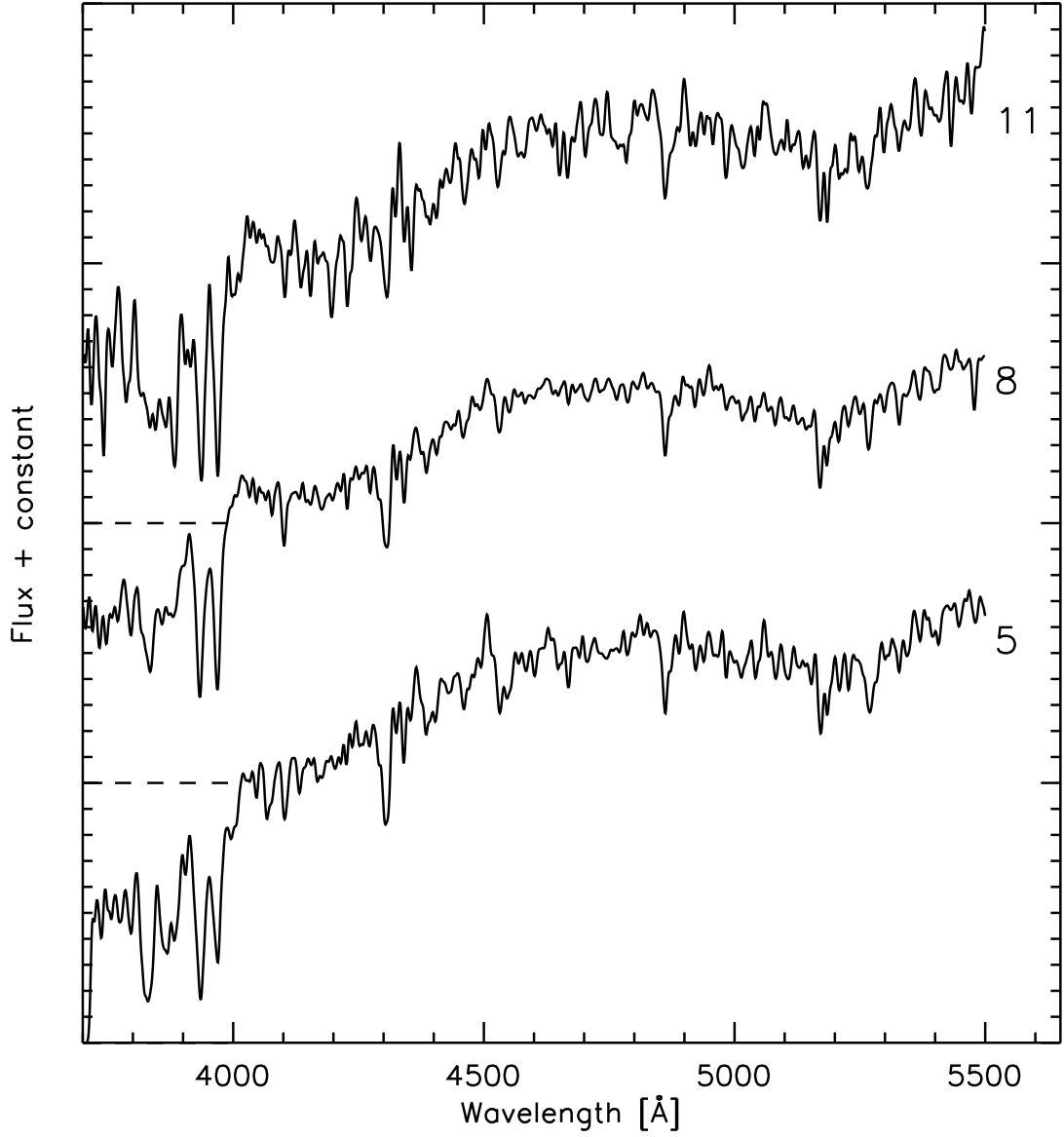}
\figcaption{\label{fig:spectra}Three sample spectra, smoothed to the
  Lick/IDS resolution. The spectra have been shifted to 0 radial velocity.
  For clarity, offsets have been added to the intensity scales of
  objects 8 and 11, as indicated by the short dashed lines.
}
\end{figure}

  Initial processing of the images (bias subtraction, flatfield correction,
cosmic ray removal etc) was done with standard tools in IRAF\footnote{
  IRAF is distributed by the National Optical
  Astronomical Observatories, which are operated by the Association of
  Universities for Research in Astronomy, Inc.~under contract with the National
  Science Foundation
}. After correction for optical distortions, each individual spectrum was 
extracted using the APALL task in the SPECRED package in IRAF. Following 
wavelength calibration based on arc lamps mounted within the 
spectrograph, small zero-point corrections (1--3 \AA) were applied by 
measuring the [O{\sc i}] skylines at 5577.338\AA\ (blue spectra) and 
6300.304\AA\ 
(red spectra).  Finally, flux calibration was applied and the individual 
spectra of each object were summed. Three sample spectra are shown in
Figure~\ref{fig:spectra}.

  Basic data for the cluster candidates are listed in Table~\ref{tab:clusters}.
The $V,I$ photometry is from the HST data used in \citet{lar01}, while
the $K$ magnitudes are from the VLT/ISAAC data in P02. Radial velocities
were determined by cross-correlating the cluster spectra with those of the
radial velocity standards, using the FXCOR task in the RV package in IRAF.
The radial velocities listed in Table~\ref{tab:clusters} are an average
of the cross-correlation results for each reference star, where the errors
are estimates of the standard error on the mean based on the two measurements.
We note that there is a systematic difference of about 100 km/s between
the radial velocity measurements for each of the two stars, suggesting that
the catalogued velocity of at least one of the stars is erroneous or
that one star might have been misidentified.
If this systematic difference is taken into account, the r.m.s. difference
between the two sets of measurements is only about 13 km/s, so
the relative radial velocities are probably more
accurate than implied by the errors listed in Table~\ref{tab:clusters}.
For NGC~4365 itself, the RC3 catalog \citep{devau91} lists a radial velocity
of $1243\pm6$ km/s.  Four objects turned out to be either likely foreground
stars or have too low signal-to-noise to extract a useful spectrum.
This leaves us with 14 confirmed clusters, with a mean radial velocity of 
$1144\pm75$ km/s and a dispersion of $279\pm67$ km/s.

\section{Measuring Lick/IDS indices}

  In order to estimate ages and metallicities for the GCs we employ the 
Lick/IDS system of absorption indices \citep{wor94}, including H$\delta_A$ 
and H$\gamma_A$ from \citet{wo97}.  After correcting for the radial velocities
in Table~\ref{tab:clusters}, we re-measured the locations of several prominent
features such as the Balmer lines, G-band and Ca {\sc ii} H$+$K lines and 
found the wavelength scales to be accurate to better than 1 \AA\ in the 
relevant range (i.e.\ $\sim4000-5500$\AA).  Because our LRIS spectra have a 
higher spectral resolution (about 5\AA\ FWHM) than the original Lick/IDS 
system (8--13 \AA ), we smoothed our spectra 
to the IDS resolution with a wavelength-dependent Gaussian kernel before 
measuring the indices.  Another difference between the original Lick/IDS 
system and our LRIS spectra is the fact that 
we are working on flux-calibrated data, whereas the original Lick/IDS spectra 
on which the Lick system is based were not flux calibrated. However, it is 
well documented that differences in flux calibration procedures affect the
equivalent widths of most Lick indices only weakly, at the level
of $\la0.1$\AA\ \citep[e.g.][]{fab85,kis98,lb02}. One possible 
exception is the \mgtwo\ index, whose continuum passbands are more widely 
separated, and where offsets between instrumental systems and the Lick
standard system have often been found \citep[e.g.][]{kun00}.

\begin{figure}
\plotone{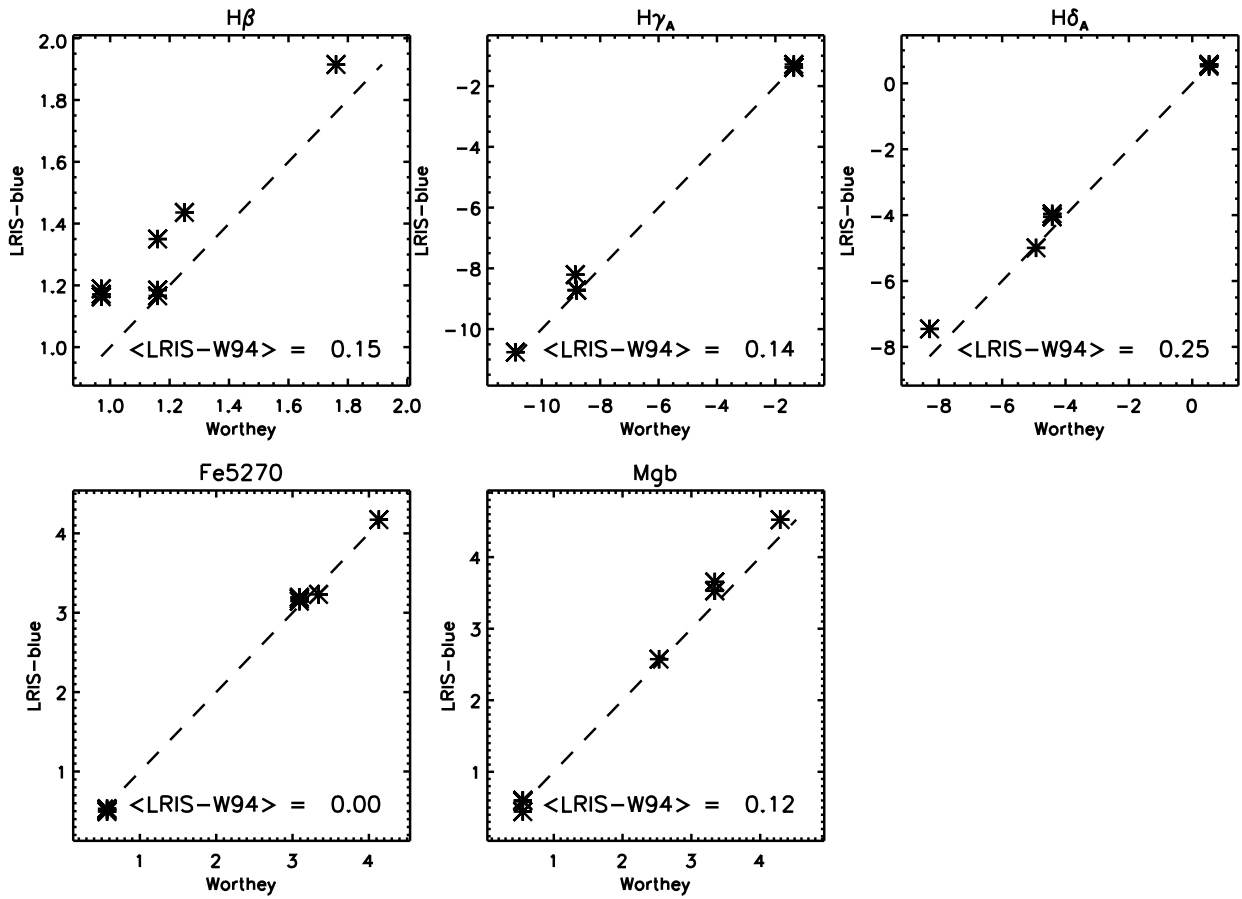}
\figcaption{\label{fig:stdcmp}Comparison of our Lick/IDS index measurements 
  for the standard stars and the \citet{wor94} standard values. 
}
\end{figure}

  For most indices we were able to check the agreement between our 
instrumental system and the Lick standard system using the
Lick/IDS standard star observations.  Figure~\ref{fig:stdcmp} shows our
standard star measurements versus the standard values 
\citep{wor94,wo97}. The mean offsets are listed in Table~\ref{tab:stdcmp}.
Even after smoothing, systematic offsets of $0.1-0.2$\AA\ remained 
between our measurements and the standard values.  Unfortunately, since 
the Lick/IDS standard stars were observed in longslit mode, they did not 
include the red continuum passbands of \mgtwo\ and Fe5335, so we were unable to 
check these two indices.  Because some of the indices fell outside the 
spectral range of the standard star observations and some of the 
globular cluster spectra have measured indices outside the range spanned by 
the standard stars, we have not attempted to correct our index 
measurements on the science spectra for the offsets between our instrumental 
system and the Lick/IDS standard system. However, we note that these small 
offsets are of no consequence to our analysis or conclusions, and are generally
smaller than the random measurement errors and model uncertainties. The 
indices used in this paper are listed in Table~\ref{tab:indices_gw}.

\section{Results}

\subsection{Ages and metallicities}
  
\begin{figure}
\plotone{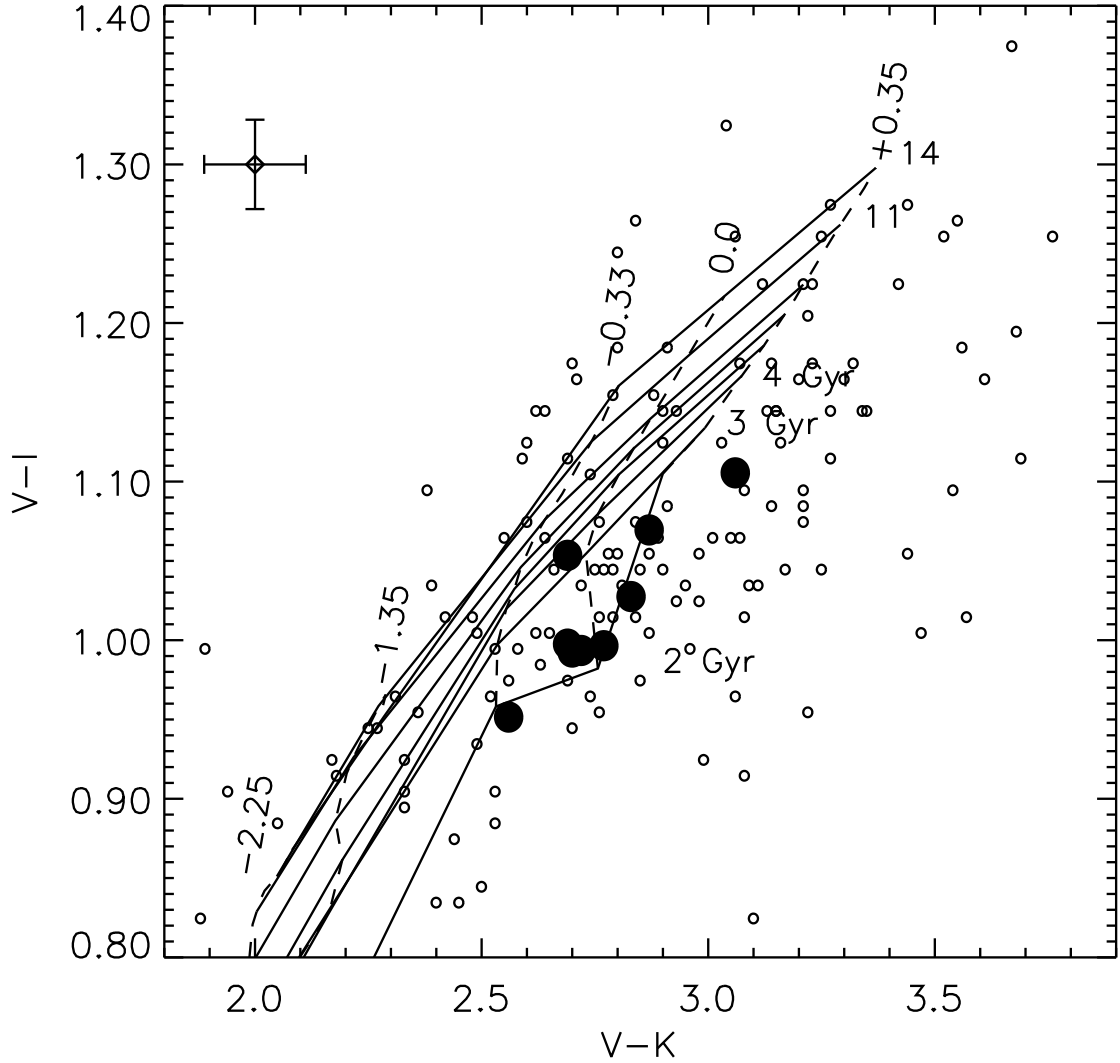}
\figcaption{\label{fig:pvk_vi}(\vk, \vi) two-color diagram for globular
  clusters in NGC~4365, shown together with a Maraston model grid. The 9 
  clusters selected for spectroscopy are shown with filled circles while 
  open circles represent the full sample from Puzia et al.  A typical error 
  bar is also shown.
}
\end{figure}

  Fig.~\ref{fig:pvk_vi} shows a (\vk, \vi) two-color diagram for GC
candidates in NGC~4365, based on the data
in P02.  The photometry has been corrected for a foreground reddening of 
$A_B = 0.091$ mag \citep{sch98}, adopting the extinction curve in 
\citet{car89}.  The 9 spectroscopically confirmed clusters with $K$-band data 
are shown with filled circles.  Also shown are single-aged stellar
population (SSP) models by \citet{mar02} for ages
of 2, 3, 4, 5, 6, 8, 11 and 14 Gyrs and metallicities of \zh = $-2.25$, 
$-1.35$, $-0.33$, 0.0 and $+0.35$. From the optical colors alone and
assuming old ages, the 9 clusters would appear to have low to intermediate 
metallicities, but their location in the two-color diagram suggests 
that they may instead be as young as 2--3 Gyrs and have metallicities near 
solar. For a more detailed discussion of the (\vk, \vi) diagram, including a 
comparison of various SSP models, we refer to P02.

\begin{figure}
\plotone{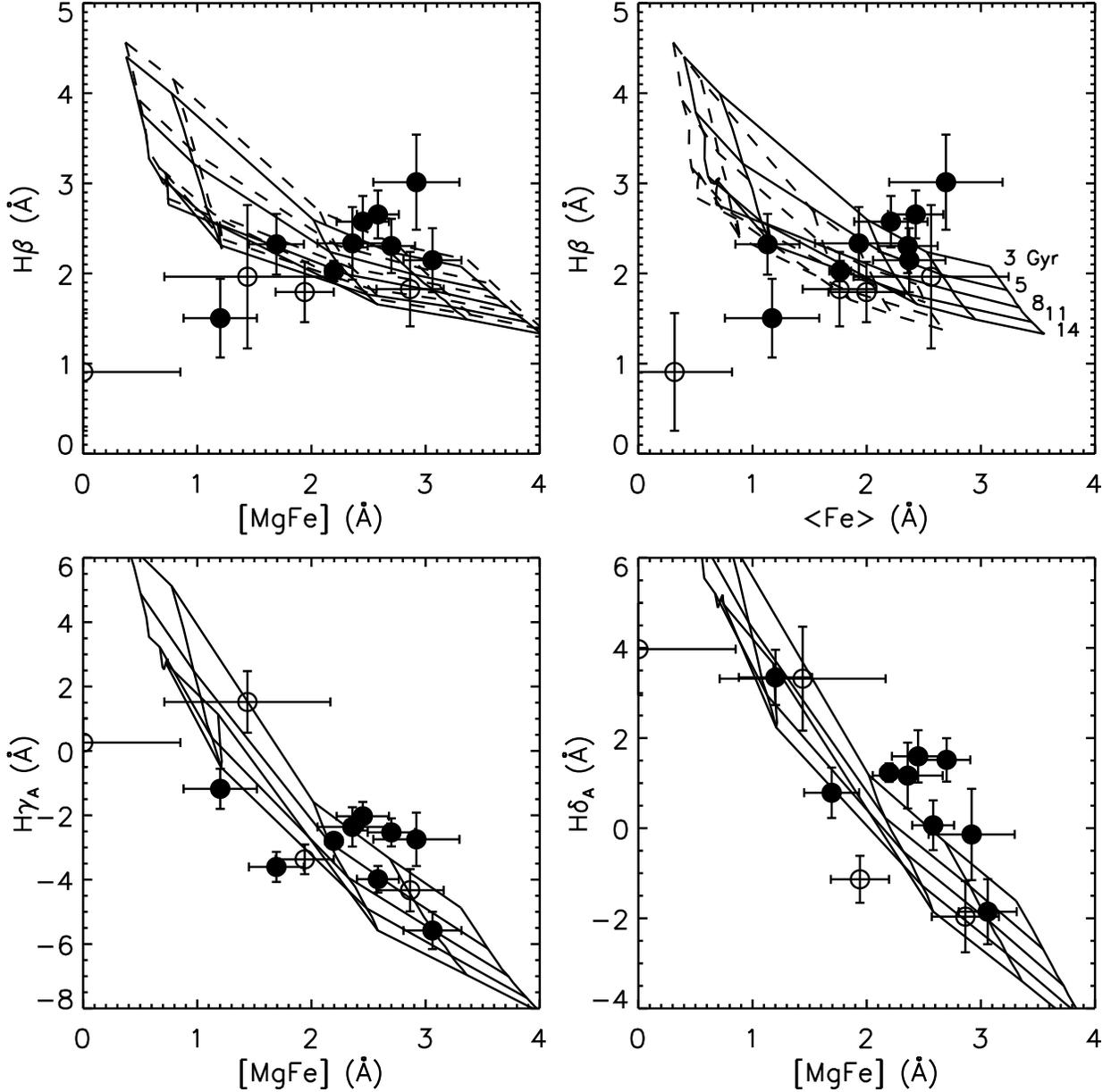}
\figcaption{\label{fig:idxs}Comparison of index measurements for
  GCs in NGC~4365 with model grids from TMB02.  The models are shown
  for ages of 3 Gyrs, 5 Gyrs, 8 Gyrs, 11 Gyrs and 14 Gyrs and
  mean metallicities \zh\ = $-2.25$, $-1.35$, $-0.33$, 0.0 and $+0.35$.
  Clusters with and without $K$-band data are plotted with filled and 
  open circles, respectively. In the top panels, solid and dashed line 
  grids indicate $\alphafe=0$ and $\alphafe=+0.5$.
}
\end{figure}

  In Fig.~\ref{fig:idxs} we show the Balmer line indices (H$\beta$, H$\gamma_A$ 
and H$\delta_A$) for the clusters vs.\ \mgfe\ and \fe , compared with SSP 
model grids by \citet[][hereafter TMB02]{tmb02}.  A variety of SSP models are 
now available in the literature, but for this work we have chosen the models by
TMB02 because they are tabulated for several different \alphafe\ values
and thus allow us to quantify the effect of varying $\alpha$-element
abundances. These models are also the first to attempt a correction for the
\alphafe\ bias in the original Lick/IDS fitting functions \citep{marea02}.
We have used the common 
definitions \fe\ = (Fe5270$+$Fe5335)/2 and \mgfe\ = $\sqrt{{\rm Mg}b\,\fe}$
\citep{gon93}.
The SSP models are shown for ages of 3, 5, 8, 11 and 14 Gyrs and the same
metallicities as in Fig.~\ref{fig:pvk_vi}.  In the plots involving 
H$\beta$, we show models with $\alphafe=0$ (solid lines) and $\alphafe=+0.5$ 
(dashed lines). In the other plots only $\alphafe=0$ models are shown.  
Clusters with and without $K$-band imaging are shown with filled and open 
circles, respectively. 
Object \#3, which has Mg$b$ $<0$, is shown at $\mgfe=0$. The blue color
of this cluster, as well as its weak Fe indices, suggest that it is very
metal-poor. It is also one of the faintest objects in our sample and 
was observed in a short ($6\arcsec$) slit, making accurate sky subtraction
difficult. The negative Mg$b$ value is most likely due to contamination by 
the [N{\sc i}] night-sky line at 5199 \AA , which is shifted into the Mg$b$
central bandpass after correcting for the radial velocity of NGC~4365.

  All four panels in Fig.~\ref{fig:idxs} confirm a spread in metallicity
with some clusters approaching Solar values, but none of the clusters
included in our small sample appear to have metallicities significantly 
above Solar.  The combined \fe\ index measures primarily Fe-peak elements 
\citep{tb95} and is therefore sensitive to \alphafe\ when plotted 
for a fixed mean metallicity \zh . The \mgfe\ grids, on the other hand, vary 
only weakly with \alphafe\ for fixed \zh , as can be seen from the
similarity between the dashed-line and solid-line grids in the upper
left panel of Fig.~\ref{fig:idxs}. Comparison of Balmer line indices 
with the models again indicates intermediate ages for many 
of the metal-rich clusters, but note that some of the filled circles (all of 
which, by selection, appear young in Fig.~\ref{fig:pvk_vi}) are actually 
better fit by the models corresponding to old ages.  Thus, there are a few
cases in which the spectroscopic and photometric data are not entirely
consistent.  Some of the clusters without IR data also appear to be 
genuinely old.  

\begin{figure}
\plotone{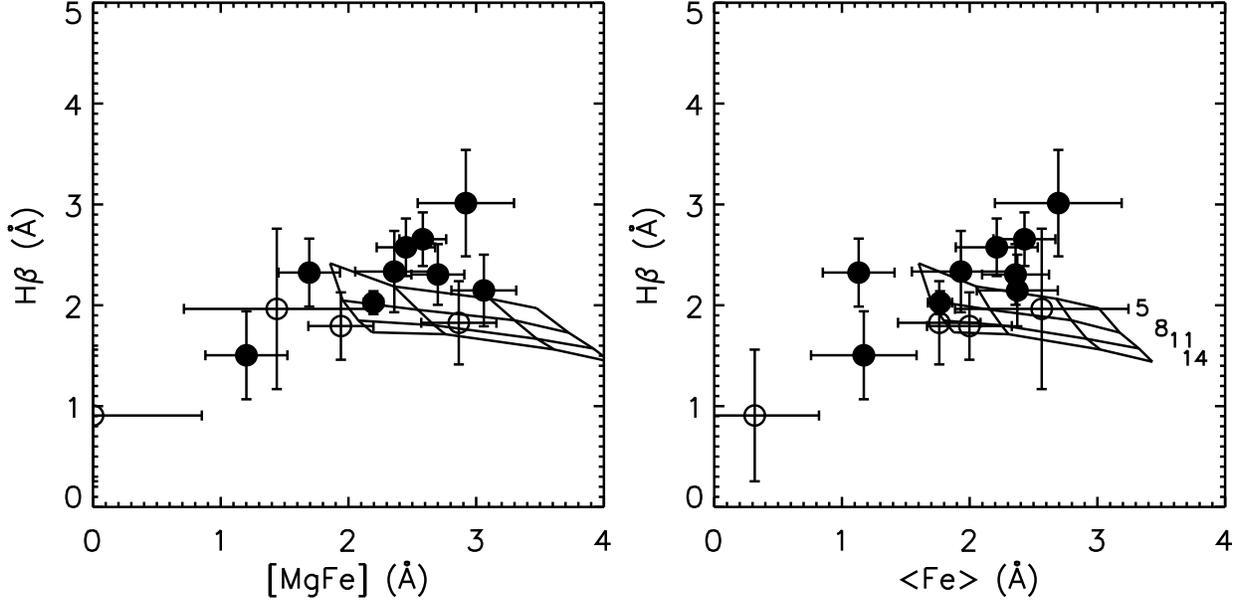}
\figcaption{\label{fig:idxs_sci}Comparison of index measurements for
  GCs in NGC~4365 with model grids from Schiavon (2002). The models
  are tabulated for ages of 5, 7.9, 11.2 and 14.1 Gyrs and
  metallicities of \zh\ = $-0.7$, $-0.4$, 0.0 and $+0.2$ --- note the
  more restricted age- and metallicity range compared to the TMB02
  models.  Symbols are the same as in Fig.~\ref{fig:idxs}.
}
\end{figure}

  The TMB02 models, like most other available SSP models, are based on the 
Lick/IDS ``fitting functions''. A new set of SSP models based on fitting 
functions derived from the spectral library by \citet{jon99}
has recently been computed by R.\ Schiavon and was kindly made available
to us (a discussion of some of the ingredients in these models is given
in \citet{schia02}). The Schiavon models cover a more restricted range in age 
and metallicity, but are useful as a comparison with the Maraston models.
In Fig.~\ref{fig:idxs_sci} we compare our measurements of H$\beta$, \mgfe\ 
and \fe\ with the Schiavon models (based on Salaris isochrones). The slope 
of the lines of constant age in the H$\beta$ vs.\ \mgfe\ and \fe\ planes 
differs somewhat 
from that in Fig.~\ref{fig:idxs}, suggesting that even relative ages based on 
Balmer line indices may still be inherently uncertain at the level of a few 
Gyrs.  Nevertheless, the picture remains clear: many of the metal-rich
clusters in NGC~4365 have stronger H$\beta$ indices than even the 5 Gyr 
models, and thus appear to be young.

  Considering the current model uncertainties and observational errors, it 
is probably not very meaningful to assign absolute ages to any individual 
clusters, but Figs.~\ref{fig:pvk_vi} -- \ref{fig:idxs_sci} consistently suggest 
that NGC~4365 hosts a substantial number of clusters as young as 2--5 Gyrs 
and with near-Solar metallicities.  However, confirmation of this result 
with a larger sample of high S/N spectra would clearly be very desirable. We 
also note that factors such as horizontal branch morphology and the luminosity 
function on the red and asymptotic giant branches may affect line indices 
and remain major uncertainties in SSP models \citep[e.g.\ ][]{schia02}.
This might account for some of the differences between the spectroscopic 
and photometric age/metallicity estimates, and could also have a significant
effect on ages derived from Balmer lines \citep{lee02}.

\subsection{$\alphafe$ abundance ratios}
\label{sec:alphafe}

\begin{figure}
\plotone{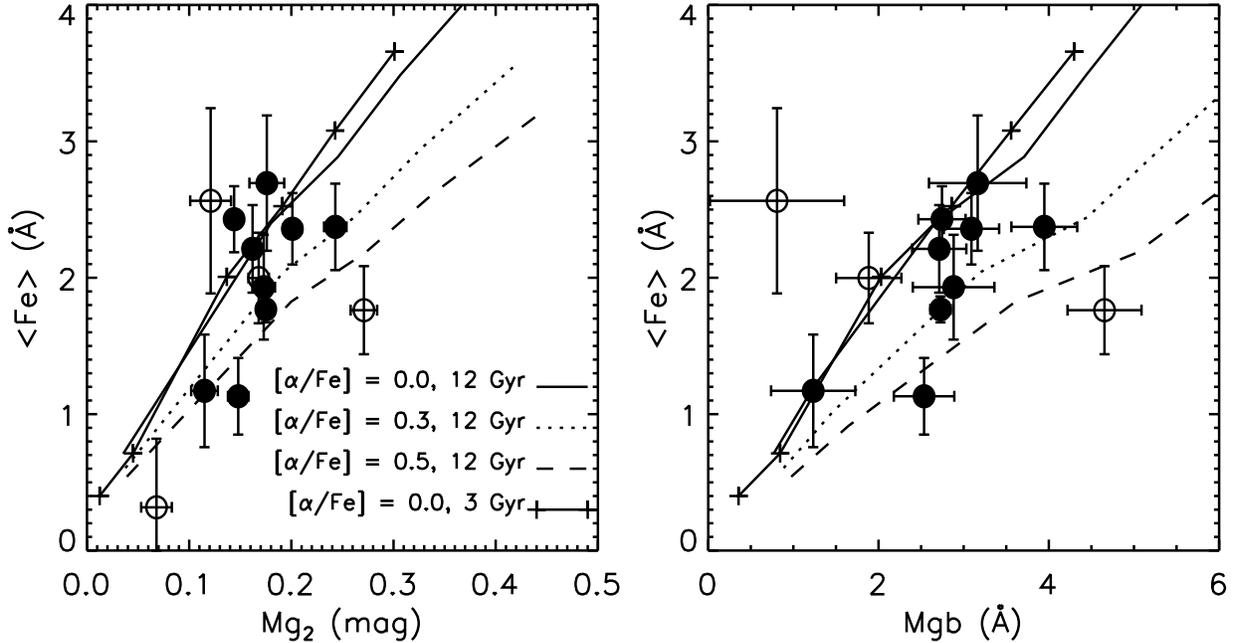}
\figcaption{\label{fig:alphafe}\fe\ vs.\ \mgtwo\ and Mg$b$, compared with 
  $\alpha$-enhanced SSP models by TMB02.
}
\end{figure}

  In Fig.~\ref{fig:alphafe} we compare our measurements of \mgtwo\ and Mg$b$
(both indicators of $\alpha$-elements) with the SSP models by TMB02.
We show the $\alphafe=0$ models for both 3 and 12 Gyrs, to illustrate 
that the relation between \mgtwo /Mg$b$ and \fe\ is only expected to depend 
weakly on age.  For the most metal-poor clusters neither 
plot provides strong constraints on \alphafe\ ratios, although the \mgtwo\ vs.\ 
\fe\ plot (which has the smallest formal errors) suggests super-solar 
\alphafe , as in other old stellar populations.  There may be a slight 
tendency for the most metal-rich clusters to have $\alphafe$ closer to solar.  
While offsets may be present between our instrumental system and the Lick 
standard system for the \mgtwo\ index, comparison of the two panels in 
Fig.~\ref{fig:alphafe} suggests that any such offset is relatively small.  
Ratios of \alphafe\ 
of $\approx+0.3$ are common in globular clusters and elliptical galaxies and 
are usually attributed to rapid early enrichment dominated by Type II SNe. For 
an intermediate-age population, one would expect the \alphafe\ ratios to be 
closer to Solar, since the gas would have been enriched with Fe-group elements 
by Type Ia SNe over several Gyrs \citep{tgb99}. Thus, the trends in 
\alphafe\ with metallicity suggested by Fig.~\ref{fig:alphafe} are again 
consistent with many of the metal-rich clusters being intermediate-age.

\section{Origin of the intermediate age population}

  The photometric optical/IR and spectroscopic data clearly hint at an 
intermediate-age, metal-rich population of GCs in NGC~4365. Although we
would expect the globular cluster populations to trace major star 
formation episodes in their host galaxy, little is currently known about
the detailed star formation history of any elliptical galaxy and
NGC~4365 is no exception.  Based on a comparison of
H$\beta$, \fe\ and \mgtwo\ measurements with a 1992 version of the Worthey SSP 
models, combined with kinematic evidence, \citet{sb95} suggested that a 
metal-rich (up to $2 \, Z_{\odot}$) and $\sim7$ Gyr old stellar population 
with a mass of $(8.1\pm2.5)\times10^9\, \msun$ exists in the central regions 
of NGC~4365.  However, new results from integral field spectroscopy of NGC~4365 
\citep{dav01} and comparison with more recent SSP models indicate a 
luminosity-weighted age for the stellar population in the galaxy of about 
14 Gyrs over a continuous central region covering 0.5 effective radii.  
NGC~4365 does not show any of the usual signs of merger activity within the 
last few Gyrs ({\sc Hi} tidal tails, disturbed morphology etc.). Although it 
does have a kinematically decoupled core (KDC), \citet{dav01} found no age 
differences between the KDC and the remainder of the galaxy, and both 
components appear to have the same elevated Mg-to-Fe ratio. An intermediate
age population may still exist within the central $\sim1\arcsec$ of the 
nucleus, but its luminosity and associated mass are probably much smaller than 
required to account for the intermediate-age GCs \citep{car97}.  Even if the 
KDC indicates a merger history, it is therefore not clear that the event 
which produced the KDC is related to the origin of the intermediate-age GCs. 

  In present-day starbursts, formation of clusters is generally accompanied 
by formation of field stars, so it would be natural to expect a stellar 
population with the same age and metallicity as the GCs. Indeed, the 
metallicity of the metal-rich GCs in many systems tends to correlate with 
that of the parent galaxy \citep[e.g.][]{fbg97}, suggesting a common origin. 
If the old age of the general stellar population in NGC~4365 and the 
intermediate age of the metal-rich GCs are both confirmed, an interesting
question will be whether or not there are any field stars associated with the 
intermediate-age GCs at all. One possibility is that some intermediate-age 
field stars do exist in NGC~4365, but constitute a too small fraction of 
the total population to strongly influence the integrated light.  

  How many young stars could be ``hidden'' in NGC~4365 without significantly
affecting the integrated light? A detailed answer to this question is
beyond the scope of the present paper, but we can make some rough
estimates. First, it is important to specify the age of the young population,
because the Balmer line strengths and luminosity per unit mass both increase
strongly at younger ages, causing even a small number of very young stars
to have a strong impact on the integrated spectrum. In a realistic modelling,
the effects of different metallicities also need to be considered. However,
here we are mainly interested in the \emph{change} in Balmer line strengths 
in the presence of young stars and for this purpose the metallicity effects 
can, at least as a first approximation, be considered second-order.  In the
following we will simply assume Solar metallicity.
The Maraston SSP models that we have used in previous sections do not tabulate 
mass-to-light ratios, so for this purpose we use models by Bruzual \& Charlot 
(2000; priv.\ comm.).  These models give M/L ratios for various broad-band 
filters, of which we will use the $B$-band numbers as the best approximation 
to the region around H$\beta$.  For 2-Gyr and 5-Gyr populations, the $B$-band 
luminosities per unit mass are about 9.5 and 3.3 times higher than at 15 Gyrs, 
respectively, assuming a Miller-Scalo IMF.  For solar metallicity, the 
TMB02 models predict an H$\beta$ equivalent width of 2.91\AA, 2.16\AA\ and 
1.51\AA\ at 2 Gyrs, 5 Gyrs and 15 Gyrs, respectively.  

  We can now estimate the H$\beta$ equivalent width for any mix of the 2 Gyr, 
5 Gyr and 15 Gyr populations by properly weighting each component. For a 15 
Gyr population mixed with a 5 Gyr population constituting 5\% of the total 
mass, the H$\beta$ EW is 1.61\AA , about 0.1\AA\ more than for a pure 15 Gyr 
population. According to the TMB02 models, this corresponds to an age of 13 
Gyrs instead of 15 Gyrs inferred from the integrated spectrum.  If the 
15 Gyr population is mixed with a 2 Gyr population, again constituting 5\% of 
the mass, the H$\beta$ EW would increase to 1.98\AA , corresponding to an
age of 7 Gyrs instead of 15 Gyrs.  In order to limit the increase in H$\beta$ 
to 0.1\AA, only about 1\% of the mass could be in the 2 Gyr population. 
For the other Balmer lines we find very similar results.

  While in reality the effects of different metallicities, IMFs, changes in 
integrated colors etc.\ need to be investigated in a more rigorous way, it is 
clear that only a small number of younger stars could be present in NGC~4365 
without noticable effects on the integrated spectrum.  The above considerations 
suggest that a mass fraction of 1\%--5\% in a 2--5 Gyr population would lead 
to a decrease of about 2 Gyrs in the luminosity-weighted age estimates (based 
on Balmer lines), and thus a mass fraction of this order could plausibly remain 
undetected. It is important to point out, however, that this calculation 
applies to the integrated light at any given position within the galaxy --- 
if, for example, the young population were more centrally concentrated, its 
relative contribution to the mass would have to remain below a few percent 
even within the central regions, and thus constitute a much smaller fraction 
of the total mass globally.

  With current data, the relative numbers of intermediate-age and old 
\emph{clusters} are poorly constrained, but P02 estimated that about 150 
clusters in NGC~4365 [out of a total $\sim2500$; \citet{az98}] might belong 
to the younger population.  This corresponds to about 6\% of the clusters 
being young.  While the formation efficiency of clusters relative to field 
stars might vary for the different populations, it is conceivable that any 
field stars associated with the intermediate-age clusters could be ``hidden'',
with the requirement that these stars are uniformly distributed throughout 
the galaxy.  P02 also estimated that 40\% -- 80\% of the clusters in their
\emph{observed} sample (i.e.\ within a $2\farcm5\times2\farcm5$ field near
the center) are young,  but because of their color-dependent detection limit,
they estimate a biasing factor of 1.8 in the detected numbers of metal-rich
(IR-bright) vs.\ metal-poor clusters. However, even a mass fraction as high 
as $\sim25\%$ in 2--5 Gyr old field stars 
in the central regions would completely dominate the integrated spectrum, and 
is clearly incompatible with a uniformly old luminosity-weighted age.
The relative numbers of globular clusters
and field stars at different metallicities have only been examined in
detail for one large elliptical, the nearby NGC~5128 \citep{hh02}, 
where the number of clusters relative to field stars seems to
\emph{decrease} with increasing metallicity. This trend, then, is 
the opposite of what would be required in NGC~4365.  Both the GC system of 
NGC~4365 and the field star population clearly warrant further study.

\section{Summary}
  
  We have presented spectroscopy for 14 globular clusters in NGC~4365, of
which 9 were suspected to be intermediate-age, metal-rich objects based on
$K$-band and optical photometry. Our spectroscopy supports this suspicion
for most of the objects and comparison with current simple stellar
population models suggests ages of 2--5 Gyrs and metallicities in the range 
$-0.4 \la \zh\ \la 0$. However, some old ($\sim$10--15 Gyrs) clusters are 
also present. Our spectra thus appear to support the idea that the apparent 
unimodal optical GC color distribution in NGC~4365 is due to a particular 
combination of GC ages and metallicities.  We see hints of a decrease in 
$\alphafe$ abundance ratios with increasing metallicity, with close to Solar 
$\alphafe$ for the metal-rich, intermediate-age clusters.  The results on GC 
ages are puzzling in the context of recent data for NGC~4365 itself, which 
suggest that the stellar population is uniformly old with an age of $\sim14$ 
Gyrs.  At most a few percent of the mass could be in a 2--5 Gyr stellar 
population without strongly affecting the integrated spectrum.  It would be
highly desirable with spectroscopy and/or IR imaging of many more 
clusters in NGC~4365 in order to better constrain the relative 
numbers of intermediate-age and old clusters.  If confirmed, 
these findings would pose a significant challenge for understanding the origin 
of GC sub-populations.

\acknowledgments

  This work was supported by  National Science Foundation grants number 
AST9900732 and AST0206139.  We are grateful to C.\ Maraston and D.\ Thomas 
for providing us with their SSP models, and to an anonymous referee for 
helpful comments.

\newpage

\begin{deluxetable}{cccccccc}
\tabletypesize{\footnotesize}
\tablewidth{0pt}
\tablecaption{ \label{tab:clusters}Globular cluster candidates in NGC~4365.
}

\tablecomments{
  No correction for Galactic foreground reddening has been applied to the
  photometry in this table.
  Coordinates are measured on the WFPC2 images.
}
\tablehead{
   ID      & RA   & DEC  & $V$     & \vi\  & $K$  & RV  &   GC?  \\
           & \multicolumn{2}{c}{J2000.0} & \multicolumn{2}{c}{HST/WFPC2} &
	     VLT/ISAAC & km s$^{-1}$ & }
\startdata
 1 & 12:24:24.083 & 7:17:44.31 & $21.98\pm0.01$ & $1.090\pm0.010$ &   \ldots        & $1265\pm52$ & Yes \\
 2 & 12:24:25.470 & 7:17:54.55 & $22.38\pm0.02$ & $0.898\pm0.027$ &   \ldots        & $1428\pm56$ & Yes \\
 3 & 12:24:25.672 & 7:18:01.30 & $22.20\pm0.02$ & $0.892\pm0.024$ &   \ldots        & $1177\pm62$ & Yes \\
 4 & 12:24:24.311 & 7:18:25.50 & $22.06\pm0.01$ & $0.987\pm0.020$ & $19.44\pm0.07$ & $1248\pm58$ & Yes \\
 5 & 12:24:25.298 & 7:18:25.61 & $21.27\pm0.01$ & $1.089\pm0.011$ & $18.52\pm0.04$ & $1628\pm54$ & Yes \\
 6 & 12:24:24.718 & 7:18:58.99 & $22.48\pm0.02$ & $0.582\pm0.036$ &   \ldots        &  $136\pm58$ & No \\
 7 & 12:24:26.654 & 7:18:55.14 & $22.02\pm0.02$ & $1.015\pm0.029$ & $18.88\pm0.05$ & $362\pm228$ & ? \\
 8 & 12:24:26.636 & 7:19:09.36 & $20.13\pm0.01$ & $1.029\pm0.006$ & $17.35\pm0.02$ & $1001\pm19$ & Yes \\
 9 & 12:24:25.914 & 7:19:34.58 & $21.57\pm0.01$ & $1.063\pm0.013$ & $18.68\pm0.04$ &  $875\pm56$ & Yes \\
10 & 12:24:29.465 & 7:19:03.49 & $22.62\pm0.02$ & $0.701\pm0.038$ &   \ldots        &  \ldots     & ? \\
11 & 12:24:27.425 & 7:19:42.22 & $22.17\pm0.02$ & $1.105\pm0.024$ & $19.24\pm0.06$ &  $514\pm49$ & Yes \\
12 & 12:24:28.724 & 7:19:38.26 & $21.45\pm0.01$ & $1.028\pm0.016$ & $18.69\pm0.04$ & $1302\pm28$ & Yes \\
13 & 12:24:29.409 & 7:19:50.07 & $21.60\pm0.01$ & $1.032\pm0.016$ & $18.77\pm0.04$ & $1115\pm43$ & Yes \\
14 & 12:24:28.595 & 7:20:08.39 & $21.63\pm0.01$ & $1.141\pm0.013$ & $18.51\pm0.04$ & $1395\pm40$ & Yes \\
15 & 12:24:30.330 & 7:20:04.26 & $21.64\pm0.01$ & $1.033\pm0.015$ & $18.89\pm0.05$ & $1003\pm35$ & Yes \\
16 & 12:24:29.409 & 7:21:35.16 & $21.33\pm0.01$ & $1.128\pm0.009$ &   \ldots        & $1188\pm40$ & Yes \\
17 & 12:24:28.869 & 7:21:57.34 & $20.56\pm0.01$ & $1.185\pm0.006$ &   \ldots        &  $880\pm48$ & Yes \\ 
18 & 12:24:31.509 & 7:22:32.15 & $21.59\pm0.01$ & $0.825\pm0.013$ &   \ldots        &  $191\pm31$ & No \\
\enddata
\end{deluxetable}

\begin{deluxetable}{lc}
\tablewidth{0pt}
\tablecaption{ \label{tab:stdcmp} Offsets between our standard star 
  measurements and values tabulated by \citet{wor94} and \citet{wo97}.}
\tablehead{Index   &   $\langle$Measured$-$Standard$\rangle$ (\AA )}
\startdata
H$\beta$    &      $0.15\pm0.03$ \\
H$\gamma_A$ &      $0.14\pm0.07$ \\
H$\delta_A$ &      $0.25\pm0.11$ \\
Fe5270      &      $0.00\pm0.03$ \\
Mg$b$       &      $0.12\pm0.05$ \\
\enddata
\end{deluxetable}

\begin{deluxetable}{lrrrrrrrrrr}
\rotate
\tabletypesize{\footnotesize}
\tablewidth{0pt}
\tablecaption{ \label{tab:indices_gw}Lick/IDS indices for globular clusters
  in NGC~4365.}
\tablehead{ ID      & H$\beta$ & H$\gamma_A$ & H$\delta_A$ & Fe5270 & Fe5335 &
                      \mgtwo\ & Mg$b$ & \fe\ & \mgfe \\
		    &  \AA     &   \AA       &   \AA       &   \AA  &  \AA   &
		      Mag &  \AA  &  \AA &  \AA  }
\startdata
  1  & $ 1.83\pm0.41$ & $-4.33\pm0.66$ & $-1.96\pm0.79$ & $ 1.67\pm0.43$ & $ 1.85\pm0.48$ & $0.271\pm0.013$ & $ 4.65\pm0.43$ & $ 1.76\pm0.32$ & $ 2.86\pm0.29$ \\
  2  & $ 1.96\pm0.80$ & $ 1.52\pm0.96$ & $ 3.32\pm1.15$ & $ 2.94\pm0.88$ & $ 2.19\pm1.04$ & $0.121\pm0.020$ & $ 0.81\pm0.79$ & $ 2.56\pm0.68$ & $ 1.44\pm0.73$ \\
  3  & $ 0.91\pm0.65$ & $ 0.26\pm0.76$ & $ 3.97\pm1.02$ & $ 1.13\pm0.65$ & $-0.49\pm0.77$ & $0.068\pm0.015$ & $-1.77\pm0.69$ & $ 0.32\pm0.50$ & $ 0.00\pm0.85$ \\
  4  & $ 1.50\pm0.44$ & $-1.18\pm0.62$ & $ 3.35\pm0.61$ & $ 1.72\pm0.53$ & $ 0.63\pm0.63$ & $0.115\pm0.013$ & $ 1.23\pm0.50$ & $ 1.17\pm0.41$ & $ 1.20\pm0.32$ \\
  5  & $ 2.65\pm0.27$ & $-3.99\pm0.41$ & $ 0.07\pm0.55$ & $ 3.12\pm0.30$ & $ 1.74\pm0.38$ & $0.144\pm0.008$ & $ 2.75\pm0.28$ & $ 2.43\pm0.24$ & $ 2.58\pm0.18$ \\
  8  & $ 2.03\pm0.11$ & $-2.79\pm0.17$ & $ 1.23\pm0.20$ & $ 1.90\pm0.12$ & $ 1.64\pm0.15$ & $0.175\pm0.003$ & $ 2.73\pm0.12$ & $ 1.77\pm0.09$ & $ 2.20\pm0.08$ \\
  9  & $ 2.58\pm0.29$ & $-2.03\pm0.44$ & $ 1.59\pm0.58$ & $ 2.18\pm0.38$ & $ 2.24\pm0.52$ & $0.162\pm0.009$ & $ 2.71\pm0.32$ & $ 2.21\pm0.32$ & $ 2.45\pm0.23$ \\
 11  & $ 3.01\pm0.53$ & $-2.75\pm0.83$ & $-0.14\pm1.01$ & $ 2.72\pm0.61$ & $ 2.67\pm0.78$ & $0.176\pm0.017$ & $ 3.16\pm0.57$ & $ 2.69\pm0.50$ & $ 2.92\pm0.38$ \\
 12  & $ 2.32\pm0.34$ & $-3.60\pm0.47$ & $ 0.79\pm0.56$ & $ 0.87\pm0.35$ & $ 1.39\pm0.43$ & $0.148\pm0.010$ & $ 2.54\pm0.36$ & $ 1.13\pm0.28$ & $ 1.69\pm0.24$ \\
 13  & $ 2.33\pm0.40$ & $-2.36\pm0.61$ & $ 1.17\pm0.73$ & $ 1.95\pm0.46$ & $ 1.92\pm0.61$ & $0.173\pm0.011$ & $ 2.88\pm0.48$ & $ 1.93\pm0.38$ & $ 2.36\pm0.31$ \\
 14  & $ 2.15\pm0.36$ & $-5.58\pm0.58$ & $-1.85\pm0.72$ & $ 3.18\pm0.40$ & $ 1.57\pm0.49$ & $0.243\pm0.011$ & $ 3.95\pm0.39$ & $ 2.37\pm0.32$ & $ 3.06\pm0.25$ \\
 15  & $ 2.31\pm0.30$ & $-2.53\pm0.44$ & $ 1.52\pm0.48$ & $ 2.91\pm0.31$ & $ 1.80\pm0.42$ & $0.201\pm0.009$ & $ 3.09\pm0.33$ & $ 2.36\pm0.26$ & $ 2.70\pm0.21$ \\
 16  & $ 1.79\pm0.33$ & $-3.37\pm0.46$ & $-1.14\pm0.52$ & $ 2.74\pm0.37$ & $ 1.25\pm0.55$ & $0.168\pm0.010$ & $ 1.89\pm0.38$ & $ 2.00\pm0.33$ & $ 1.94\pm0.25$ \\
 17  & $ 1.54\pm0.17$ & $-4.56\pm0.25$ & $-0.29\pm0.32$ & \ldots  & \ldots  & \ldots  & \ldots  & \ldots  & \ldots  \\
\enddata
\end{deluxetable}

\newpage

\begin{thebibliography}{}
\bibitem[Ashman \& Zepf(1998)]{az98}
  Ashman, K. M. \& Zepf, S. E. 1998, 
  ``Globular Cluster Systems'', Cambridge Astrophysics Series, Cambridge
  University Press
\bibitem[Barbier-Brossat \& Figon(2000)]{bar00}
  Barbier-Brossat M. \& Figon P. 2000, \aaps, 142, 217
\bibitem[Cardelli, Clayton and Mathis(1989)]{car89}
  Cardelli, J. A., Clayton, G. C. and Mathis, J. S. 1989, \apj, 345, 245
\bibitem[Carollo et al.(1997)]{car97} 
  Carollo, C.~M., Franx, M., Illingworth, G.~D., \& Forbes, D.~A.\ 
  1997, \apj, 481, 710 
\bibitem[Davies et al.(2001)]{dav01} 
  Davies, R.~L.~et al.\ 2001, \apjl, 548, L33 
\bibitem[Faber et al.(1985)]{fab85}
  Faber, S.\ M., Friel, E.\ D., Burstein, D., \& Gaskell, C.\ M.\ 1985,
  \apjs, 57, 711
\bibitem[Forbes, Brodie, \& Grillmair(1997)]{fbg97} 
  Forbes, D.~A., Brodie, J.~P., \& Grillmair, C.~J.\ 1997, \aj, 113, 1652 
\bibitem[Forbes et al.(1996)]{for96} 
  Forbes, D.~A., Franx, M., Illingworth, G.~D., \& Carollo, C.~M.\ 1996, 
  \apj, 467, 126 
\bibitem[Gebhardt \& Kissler-Patig(1999)]{gk99} 
  Gebhardt, K.~\& Kissler-Patig, M.\ 1999, \aj, 118, 1526 
\bibitem[Gonz{\'a}lez(1993)]{gon93}
  Gonz{\'a}lez, J.\ J., 1993, PhD Thesis, Univ.\ of California, Santa Cruz
\bibitem[Goudfrooij et al.(2001)]{goud01} 
  Goudfrooij, P., Alonso, M.~V., Maraston, C., \& Minniti, D.\ 2001, 
  \mnras, 328, 237 
\bibitem[Harris \& Harris(2002)]{hh02} 
  Harris, W.~E.~\& Harris, G.~L.~H.\ 2002, \aj, 123, 3108 
\bibitem[Jones(1999)]{jon99}
  Jones, L. A. 1999, PhD Thesis, University of North Carolina
\bibitem[Kissler-Patig et al.(1998)]{kis98} 
  Kissler-Patig, M., Brodie, J.~P., Schroder, L.~L., Forbes, D.~A., 
  Grillmair, C.~J., \& Huchra, J.~P.\ 1998, \aj, 115, 105 
\bibitem[Kundu \& Whitmore(2001)]{kw01} 
  Kundu, A.~\& Whitmore, B.~C.\ 2001, \aj, 121, 2950 
\bibitem[Kuntschner \& Davies(1998)]{kd98} 
  Kuntschner, H.~\& Davies, R.~L.\ 1998, \mnras, 295, L29 
\bibitem[Kuntschner(2000)]{kun00}
  Kuntschner, H.\ 2000, \mnras, 315, 184
\bibitem[Larsen et al.(2001)]{lar01} 
  Larsen, S.~S., Brodie, J.~P., Huchra, J.~P., Forbes, D.~A.\ and
  Grillmair, C.\ 2001, \aj, 121, 2974
\bibitem[Larsen \& Brodie(2002)]{lb02}
  Larsen, S.\ S.\ \& Brodie, J.\ P., 2002, \aj, 123, 1488
\bibitem[Lee et al.(2002)]{lee02}
  Lee, H.-C., Lee, Y.-W. and Gibson, B.~K., 2002, in:
  ``Extragalactic Globular Cluster Systens'', Sprinter-Verlag, ed.\
  M.\ Kissler-Patig
\bibitem[Maraston(2002)]{mar02}
  Maraston, C.\ 2002, in preparation
\bibitem[Maraston et al.(2002)]{marea02}
  Maraston, C., Greggio, L., Renzini, A., Ortolani, S., Saglia, R.~P., 
  Puzia, T.~H., \& Kissler-Patig, M.\ 2002, A\&A, submitted
  (astro-ph/0209220)
\bibitem[Massey et al.(1988)]{mas88} 
  Massey, P., Strobel, K., Barnes, J.~V., \& Anderson, E.\ 1988, \apj, 328, 315 
\bibitem[Oke et al.(1995)]{oke95}
  Oke, J.\ B.\, Cohen, J.\ G., Carr, M.\ et al.\ 1995, \pasp, 107, 375
\bibitem[Puzia et al.(2002)]{puz02} 
  Puzia, T.~H., Zepf, S.~E., Kissler-Patig, M., Hilker, M., Minniti, D., 
  \& Goudfrooij, P.\ 2002, A\&A 391, 453
\bibitem[Schiavon et al.(2002)]{schia02}
  Schiavon, R.\ P., Faber, S.\ M., Rose, J.\ A., Castilho, B.\ V., 2002, \aj,
  submitted (astro-ph/0109365)
\bibitem[Schlegel et al.(1998)]{sch98}
  Schlegel, D. J., Finkbeiner, D. P., and Davis, M. 1998, \apj, 500, 525
\bibitem[Surma \& Bender(1995)]{sb95}
  Surma, P. \& Bender, R. 1995, \aap, 298, 405 
\bibitem[Thomas, Greggio, \& Bender(1999)]{tgb99} 
  Thomas, D., Greggio, L., \& Bender, R.\ 1999, \mnras, 302, 537 
\bibitem[Thomas, Maraston, \& Bender(2002)]{tmb02} 
  Thomas, D., Maraston, C., \& Bender, R.\ 2002, MNRAS, submitted
  (astro-ph/0209250)
\bibitem[Tripicco \& Bell(1995)]{tb95}
  Tripicco, M.~J.~\& Bell, R.~A.\ 1995, \aj, 110, 3035
\bibitem[de Vaucouleurs et al.(1991)]{devau91}
  de Vaucouleurs, G., de Vaucouleurs, A., Corwin, H.~G.,
  Buta, R.~J., Paturel, G., Fouqu{\'e}, P.\ 1991,
  ``Third Reference Catalogue of Bright Galaxies'',
  Springer-Verlag New York
\bibitem[Worthey et al.(1994)]{wor94}
  Worthey, G., Faber, S.\ M., Gonz{\'a}lez, J.\ J., and Burstein, D.\ 1994,
  \apjs, 94, 687
\bibitem[Worthey \& Ottaviani(1997)]{wo97}
  Worthey, G.~\& Ottaviani, D.~L.\ 1997, \apjs, 111, 377
\end{thebibliography}
\end{document}